# Editorial: Techniques and Methods for Astrophysical Data Visualization

Brian R. Kent, PASP Guest Editor

National Radio Astronomy Observatory, 520 Edgemont Road, Charlottesville, VA, 22903, USA; bkent@nrao.edu

The tools of observational and theoretical astrophysics continue to change and evolve as new scientific questions need to be answered, bringing with those changes a number of technological challenges. In response to those challenges, the toolkit of the astrophysicist is also changing. How do we explore our data in modern astrophysics? How can we leverage the exciting technologies that lie untapped at our fingertips? What scientific discoveries lie dormant, waiting to be found in new parameter spaces of existing data? The design of observations, surveys, and theoretical simulations coupled with traditional scientific analysis only scratches the surface of potential scientific research. Through advanced data visualization techniques, we aim to look at new and potentially unexplored domains of astrophysical data.

Astrophysics continues to be a science of the extremes - the largest, complex, and most precise machines with the fastest hardware on Earth exploring unfathomable distances in the Universe. At its core astrophysics is a physical science, its laboratory being the largest, most remote, and often the most extreme. However, the field has also emerged as a leader at the forefront of the data sciences. There are often similarities between scientific fields in terms of data visualization requirements. We continue to push the boundaries of data visualization through innovative techniques utilizing the latest advances in hardware, software, and cloud technologies. Visualization methods and techniques can serve as a complement to data analysis software, stand on their own for data exploration, or inspire with impressive visuals for science, technology, education, mathematics (STEM) and public outreach.

In this PASP Special Focus Issue, *Techniques and Methods for Astrophysical Data Visualization*, the authors explore a wide range of visualization techniques aimed at spearheading cutting edge ways of exploratory data discovery and analysis. Observational and theoretical experiments in astrophysics now operate in a mode of high-rate data acquisition. The raw and processed data sizes as well as the storage costs require innovative solutions to effectively visualize and move those data (Berriman & Groom 2011). Visualization plays an important part in the management and dissemination of observational and theoretical science data as acquisition rates of both telescopes and simulations continue to increase.

This focus issue examines visualization methods (Baines et al.; Fluke et al.; Madura; Rector et al. 2017, Diemer et al. 2017), software libraries for astronomy and analysis (Berriman et al.; Muna 2017), software implementations from other fields (Gárate; Kent; Naiman et al.; Taylor 2017), and science results from new visualization techniques (Argudo-Fernández et al.; Vogt et al.; Pomarède et al. 2017). The papers by Berriman, Gárate, Naiman, Kent, Pomarède, and Taylor feature video media content. Articles by Vogt and Madura feature interactive 3D content.

I would like to thank all of the authors for their contributions to this PASP special focus issue as well as Editor-in-chief Jeff Mangum, Elizabeth Ellor, Tim Bauer, and the PASP Publishing Team for their support in the publication of this issue. The papers have laid an important groundwork for the years to come as we enter a new era of data visualization, exploration and scientific discovery. Many of the submissions have useful tools, tutorials and source code to aid in research. The authors and I invite the astronomical community to explore their data using these tools and adapt them for use in their own specialized research.

**References**

Argudo-Fernández, M. et al. 2017, PASP, this issue
Baines, D., et al. 2017, PASP, this issue
Berriman, B. & Groom, S. 2011, ACM Queue, 9, 10
Berriman, B. et al. 2017, PASP, this issue
Diemer, B. et al. 2017, PASP, this issue
Fluke, C., et al. 2017, PASP, this issue
Gárate, M. 2017, PASP, this issue
Kent, B. R. 2017, PASP, this issue
Madura, T. 2017, PASP, this issue
Muna, D., 2017, PASP, this issue
Naiman, J. et al. 2017, PASP, this issue
Pomarède, D. et al. 2017, PASP, this issue
Rector, T., et al. 2017, PASP, this issue
Taylor, R. 2017, PASP, this issue
Vogt, F., et al. 2017, PASP, this issue

The titles below link to their corresponding PASP pages.  The arXiv links lead to astro-ph postings.

**Journal Special Focus Issue Page:**
http://iopscience.iop.org/journal/1538-3873/page/Techniques-and-Methods-for-Astrophysical-Data-Visualization

*Focus Issue Video Abstract:*
https://www.youtube.com/watch?v=AACsyyr_NZs

Editorial: Techniques and Methods for Astrophysical Data Visualization
Brian R. Kent 2017 *PASP* **129** 058001

Visualization of Multi-mission Astronomical Data with ESASky
Deborah Baines *et al.* 2017 *PASP* **129** 028001          https://arxiv.org/abs/1701.02533

Visualizing Three-dimensional Volumetric Data with an Arbitrary Coordinate System
R. Taylor 2017 *PASP* **129** 028002          https://arxiv.org/abs/1611.02517

Cosmography and Data Visualization
Daniel Pomarède *et al.* 2017 *PASP* **129** 058002          https://arxiv.org/abs/1702.01941

Introducing Nightlight: A New FITS Viewer
Demitri Muna 2017 *PASP* **129** 058003

Spherical Panoramas for Astrophysical Data Visualization
Brian R. Kent 2017 *PASP* **129** 058004          https://arxiv.org/abs/1701.08807

LSSGalPy: Interactive Visualization of the Large-scale Environment Around Galaxies
M. Argudo-Fernández *et al.* 2017 *PASP* **129** 058005          https://arxiv.org/abs/1702.04268

The Application of the Montage Image Mosaic Engine To The Visualization Of Astronomical Images
G. Bruce Berriman and J. C. Good 2017 *PASP* **129** 058006          https://arxiv.org/abs/1702.02593

The Aesthetics of Astrophysics: How to Make Appealing Color-composite Images that Convey the Science
Travis A. Rector *et al.* 2017 *PASP* **129** 058007          https://arxiv.org/abs/1703.00490

Houdini for Astrophysical Visualization
J. P. Naiman *et al.* 2017 *PASP* **129** 058008          https://arxiv.org/abs/1701.01730

Sports Stars: Analyzing the Performance of Astronomers at Visualization-based Discovery
C. J. Fluke *et al.* 2017 *PASP* **129** 058009          https://arxiv.org/abs/1702.04829

Voxel Datacubes for 3D Visualization in Blender
Matías Gárate 2017 *PASP* **129** 058010          https://arxiv.org/abs/1611.06965

A Case Study in Astronomical 3D Printing: The Mysterious η Carinae
Thomas I. Madura 2017 *PASP* **129** 058011          https://arxiv.org/abs/1611.09994

Linking the X3D Pathway to Integral Field Spectrographs: YSNR 1E 0102.2-7219 in the SMC as a Case Study
Frédéric P. A. Vogt *et al.* 2017 *PASP* **129** 058012          https://arxiv.org/abs/1611.03862

The Fabric of the Universe: Exploring the Cosmic Web in 3D Prints and Woven Textiles
Benedikt Diemer and Isaac Facio 2017 *PASP* **129** 058013          https://arxiv.org/abs/1702.03897